\documentclass[aps, twocolumn, showpacs, amsmath]{revtex4}
\usepackage{dcolumn}
\usepackage{bm}
\usepackage{graphicx}
\usepackage{lipsum}
\usepackage{color}
\usepackage{subfigure}
\usepackage{hyperref}
\usepackage{latexsym}
\usepackage{amsthm}
\usepackage{amssymb}
\usepackage{braket}
\DeclareGraphicsExtensions{.jpg,.pdf, .mps, .png, .eps, .ps, .EPS,.gif}

\begin{document}
\def\be{\begin{equation}}
\def\ee{\end{equation}}

\def\bc{\begin{center}}
\def\ec{\end{center}}
\def\bea{\begin{eqnarray}}
\def\eea{\end{eqnarray}}
\newcommand{\avg}[1]{\langle{#1}\rangle}
\newcommand{\Avg}[1]{\left\langle{#1}\right\rangle}

\def\ie{\textit{i.\,e.,}}
\def\etal{\textit{et al.}}
\def\m{\vec{m}}
\def\G{\mathcal{G}}

\newcommand{\davide}[1]{{\bf\color{blue}#1}}
\newcommand{\gin}[1]{{\bf\color{green}#1}}
\newcommand{\bob}[1]{{\bf\color{red}#1}}

\title{Island and lake size distributions in Gradient Percolation}

\author{S. S. Manna}
\affiliation{Satyendra Nath Bose National Centre for Basic Sciences, Block-JD, Sector-III, Salt Lake, Kolkata-700106, India}

\begin{abstract}
      The well known problem of gradient percolation has been revisited to study the probability 
   distribution of island sizes. It is observed that like the ordinary percolation, this distribution 
   is also described by a power law decaying function but the associated critical exponents are found to be 
   different. Because of the underlying gradient for the occupation probability, 
   the average value of the island sizes also has a gradient. The variation of the average island 
   size with the probability of occupation along the gradient has been studied together with its 
   scaling analysis. Further, we have introduced and studied the gradient bond percolation and on 
   studying the island size distribution statistics, we have obtained very similar results. We have also 
   studied the characteristics of the diffusion profile of the particle system on a lattice 
   which is initially half filled and half empty. Here also we observe the same value for the island 
   size probability distribution exponent. Finally, the same study has been repeated for the 
   nonlinear gradient percolation and the value of the island size distribution exponent is found to be a function 
   of the strength of the nonlinear parameter.
\end{abstract}

\maketitle

\section{Introduction}
\label{sec:introduction}

      When the atoms of a solid material of type A diffuses into another solid material of type B,
   and vice versa, the properties of the mixing region have been found to be very 
   interesting. For example, when the junction of two conducting materials 
   are heated to make the electrical contact the diffusion of both types of 
   conducting atoms takes place. 
 
      During early 1980s the statistical properties of the mixing region had
   attracted a lot of research interests \cite {Sapoval}. In particular, studying the structure 
   and properties of the interface of either type of solids had received much 
   attention. There were many reasons for that. For example, it had been found 
   that the interface has a fractal structure up to the diffusion length scale \cite {Rosso,Rosso1}. 
   Secondly, the interface sites have a Gaussian density profile. The dispersion 
   of the Gaussian profile grows with time \cite {Sapoval}.

      Models of the diffusion processes of one species of atoms into another species of atoms
   gave rise to the concept of gradient percolation \cite {Sapoval,Rosso,Rosso1}. Here the density gradients of A atoms (and B 
   atoms) have been created along the $+x$ (and -$x$) axis only. Let the density profile 
   of A atoms be denoted by $p(x,t)$ depending on both the space $(x)$ as well as the time 
   $(t)$ coordinates. It had been shown that the density profile follows a decaying 
   complimentary error function: $p(x,t) = \verb!erfc!(x/\ell_D)$, where $\ell_D = 
   2(Dt)^{1/2}$ is the time dependent diffusion length \cite {Dietrich}. One side of the lattice has 
   been enriched with A atoms, whereas its opposite side has the high density of B 
   atoms and the intermediate region has the mixture of A and B atoms. However, in the simulation
   process the actual diffusion of atoms have not been executed. Instead, the independent 
   positional configurations of A and B atoms have been generated using the density 
   variable $p(x,t)$ as the probability of occupation of A atoms similar to the percolation 
   problem. 

      The problem of Percolation is considered as one of the simplest but non-trivial
   models of statistical physics. This model is typically defined on a regular lattice
   as a binary state problem. Every lattice site is randomly occupied with a tunable 
   probability $p$ and left vacant with probability $(1-p)$. This process thus mimics 
   the order-disorder transition taking place at a critical value of $p_c$ \cite 
   {Broadbent,Stauffer,Grimmett,Sahimi}. It has been shown that the fractal dimension 
   $D_h$ of the `hull' or outer boundary of percolation clusters close to criticality
   is $1.74 \pm 0.02$ \cite {Voss}. 

      The model of linear gradient percolation has been defined on a two dimensional 
   substrate, e.g., a square lattice of finite size \cite {Rosso,Rosso1}. It was found 
   that instead of the actual density profile of the diffusing A atoms, a simple time 
   independent linear profile $p(x)$ for the probability of occupation with constant
   gradient keeps all static properties intact. Very interestingly, it has been found 
   that the occupation probability corresponding to the mean $x$ position of the A atoms 
   on the interface is the ordinary site percolation threshold $p_c$ of the same lattice. This information 
   had helped in estimating the value of $p_c$ more accurately. A perimeter generating random walk 
   had been devised to trace all A atoms on the interface \cite {Ziff}. Using this method
   the percolation threshold of the square lattice had been estimated very accurately 
   which is $p_c = 0.59275 \pm 0.00003$ \cite {Ziff1,Ziff2}. Further, the fractal 
   dimension $d_f$ of the interface has been estimated to be exactly equal to 
   7/4 which is the fractal dimension of the perimeter of the spanning clusters
   at the percolation thresholds of the ordinary percolation. Gradient percolation is still an active area of
   research \cite {Tencer}.

      Following the terminologies used in \cite {Rosso} we assume that an arbitrary 
   random configuration of A and B atoms of gradient percolation has both ``infinite 
   A'' cluster and ``infinite B'' cluster on two opposite sides of the lattice. 
   In addition, at the middle region of the lattice there are many small finite 
   size clusters of both types of atoms, which are called the ``islands'' of A 
   atoms within the sea of B atoms and the ``lakes'' of B atoms within the sea 
   of A atoms. 

      Question is, what may be the probability distributions of the sizes of 
   these islands and lakes? To the best of our knowledge this issue had not been 
   explored in the literature till date. Therefore, we numerically study these 
   distributions from different approaches and report the results in this paper.

      In the next section II we describe our simulations and results of gradient site
   percolation on the square lattice. Here we have studied the probability distribution
   of island sizes and their scaling analysis. Further, we have studied the variation
   of the average values of the island sizes along the direction of the probability
   gradient. In section III we have introduced the gradient bond percolation model
   and have studied the same quantities associated with island sizes in this model
   as well. Later, in section IV we have studied the relaxation of an initially dense
   system of particles using the random walk diffusion dynamics. Primarily, we have 
   focused to the probability distribution of the island sizes and observed that the
   associated exponent is different from that of the ordinary percolation in two
   dimension. Finally, the nonlinear gradient percolation has been studied in section V
   with tunable strength parameter. We summarize in section VI.
   
\section{Islands in gradient site percolation}

\begin{figure}[t]
\begin {center}
\includegraphics[width=7.0cm]{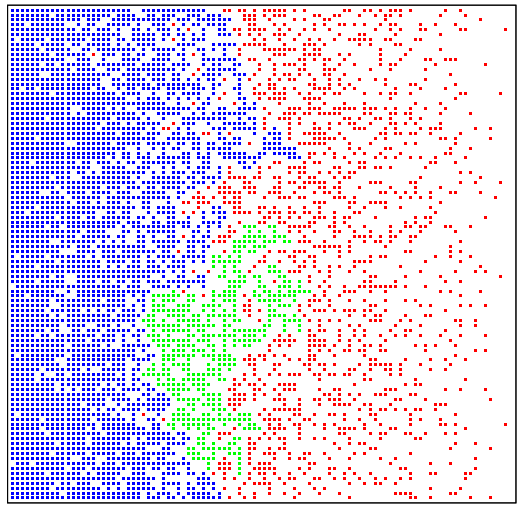}
\end {center}
\caption{
On a square lattice of size $L$ = 100, sites have been occupied with probability 
$p(x) = 1-x/(L+1)$ i.e., gradient $g = 1/(L+1)$. Marked sites represent A atoms, 
whereas, B atoms have not been shown. The infinite A cluster has 3048 sites (blue), 
the largest island has 489 sites (green) and all other islands has a total of 1381 
sites (red). The fraction of occupied sites is $\approx 0.4918$.
}
\label {FIG01}
\end{figure}

      We first identify the finite size islands. On a square lattice of
   size $L \times L$, periodic along the $y$ axis, we fill up the lattice randomly using a probability function 
   $p(x) = 1-x/(L+1)$ which has a linear gradient along the $x$-axis. All sites
   $(x,y)$ on the vertical column at a fixed $x$ have been occupied with the same
   probability $p(x)$. For the small values of $x$, the probability of occupation 
   is high and therefore most of the sites are occupied. Fig. \ref {FIG01} represents
   a typical gradient percolation configuration where all occupied sites represent 
   A atoms and all vacant sites represent B atoms. All the occupied sites which are 
   connected to the left boundary at $x = 1$ constitute the infinite A cluster and are marked 
   in blue color. As $x$ increases, the value of $p(x)$ decreases, and small lakes 
   of B sites appear within the infinite A cluster.

      On the other hand, for $x \sim L$, the value of $p(x)$ is small, so also the 
   sizes of the islands shown in red color. Clearly, there exists a gradient of the 
   island sizes as well. The average size of the islands decreases as $x$ increases. 
   How far the infinite A cluster is extended? It can be extended deep into the 
   infinite B cluster. However, the $x$ coordinates of the lattice sites on the interface 
   or on the hull of the infinite A cluster has a mean value $x_c$. It is now well known in the literature 
   that the value of $x_c$ is such that $p(x_c)=p_c$, where $p_c$ is the
   percolation threshold of the ordinary site percolation on the same lattice. The 
   interface of the infinite A cluster is defined by the sites of those A atoms which 
   have at least one B atom in the neighboring, or next neighboring site which is 
   connected via at least one path of B atoms to the right boundary at $x = L$. Therefore, $x_c$ is 
   the mean of the $x$-coordinates of all the interface sites. Most of the finite size islands 
   appear in the right side of the interface of infinite A cluster.

\begin{figure}[t]
\begin {center}
\includegraphics[width=6.0cm]{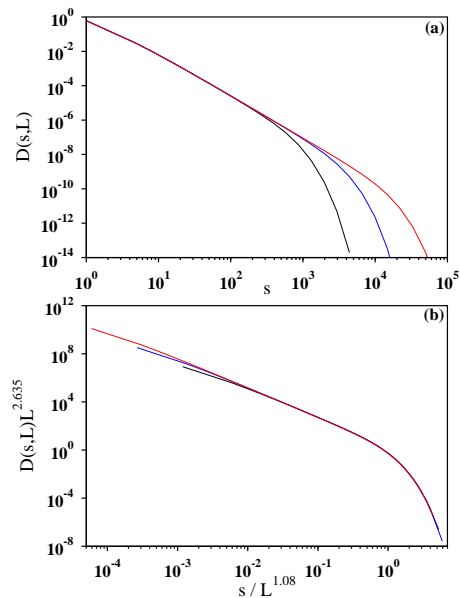}
\end {center}
\caption{
(a) The probability distribution $D(s,L)$ of the island sizes have been plotted
against the island size $s$ for square lattices of size $L \times L$. The system
size dependent linear probability gradient $g = 1/L$ have been used for $L$ = 512 
(black), 2048 (blue), and 8192 (red).
(b) The same data have been replotted after scaling the $x$ and $y$ axes using 
the scale factors $L^{1.08}$ and $L^{2.635}$ respectively. The data collapse
appears to be excellent implying the island size distribution exponent 
$\tau = 2.635/ 1.08 \approx 2.44$.}
\label {FIG02}
\end{figure}
  
      Similar to the ordinary percolation problem, we define the probability distribution $D(s,L)$
   of the island sizes $s$ of all the islands of occupied sites. Consider a typical
   configuration as shown in Fig. \ref {FIG01}. We mark different islands by different label 
   numbers (equivalent to coloring of different islands using different colors) and estimate their 
   sizes by running a simulation process called the `burning algorithm'. In this process a single 
   scan of the entire lattice enables us to identify and estimate the sizes of all distinct islands.
   The infinite A cluster is ignored and the frequency distribution of the sizes of the finite islands
   are collected in an array. This data collection process has been repeated over a large number of 
   independent configurations and the frequencies of occurrences has been cumulatively added to the 
   same array. Finally, this data has been normalized to obtain the probability distribution $D(s,L)$. 

\begin{figure}[t]
\begin {center}
\includegraphics[width=6.0cm]{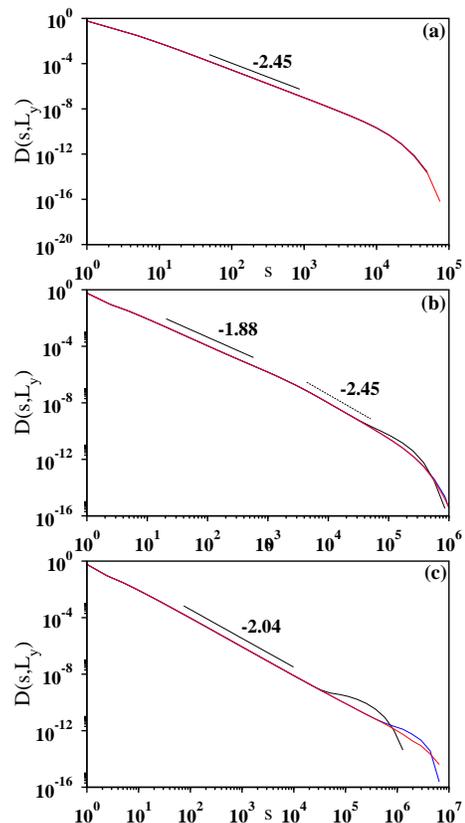}
\end {center}
\caption{
    Gradient percolation on the square lattice of rectangular shape having sides: 
    $L_x$ = 8192 and $L_y$ = 512 (black), 2048 (blue), and 8192 (red). A fixed 
    linear probability gradient $g$ has been applied along the $x$-axis.
(a) $g=2^{-13}$: The probability 
    distribution $D(s,L_y)$ of the island sizes have been plotted against the island 
    size $s$. The value of the common slope in the intermediate region is $\tau = 2.45$.
(b) $g=2^{-17}$: It is evident that there are two regimes with 
    different slopes. For small island sizes, the value of the slope is $\approx$ 1.88 (solid line), whereas, 
    in the large island regime it is $\approx$ 2.45 (dashed line).
(c) $g=2^{-21}$: A major part of the intermediate region 
    has the slope 2.04 which is close to the value of $\tau$ for ordinary percolation.
}
\label {FIG03}
\end{figure}

      In Fig. \ref {FIG02}(a), we have plotted the logarithmically binned distribution 
   data $D(s,L)$ against $s$ on a double logarithmic scale for three different system 
   sizes $L$ = 512, 2048, 8192 using the system size dependent linear probability 
   gradient $g = 1/L$. All three curves exhibit the typical signatures of the power law 
   distribution, i.e., a straight linear part in the intermediate region, followed by 
   a sharp downward bending at some characteristic cut-off island size $s_c(L) \sim 
   L^{\alpha}$ which depends on the system size. 

      We observe that the distributions $D(s,L)$ scales nicely using suitable powers of $L$, like:
\begin {equation}
D(s,L)L^{\beta} \sim {\cal G}(s/L^{\alpha})
\end {equation}
   where, ${\cal G}(x)$ is the scaling function such that ${\cal G}(x) \rightarrow x^{-\tau}$ for
   $x << 1$ and ${\cal G}(x) \rightarrow constant$ for $x >> 1$. The limiting distribution
   $D(s) = \lim_{L\to\infty}D(s,L) \sim s^{-\tau}$ must be independent of $L$ which leads to $\tau = \beta/\alpha$.
   In Fig. \ref {FIG02}(b), a finite size scaling of the same data has been 
   attempted. The $x$ and $y$ axes have been transformed by dividing the $x$ coordinates by 
   $L^{\alpha}$ and multiplying the $y$ coordinates by $L^{\beta}$. A careful fine tuning of the 
   values of the scaling exponents $\alpha$ and $\beta$ yields the best collapse of the data
   corresponding to $\alpha \approx 1.08$ and $\beta \approx 2.635$ which gives $\tau = \beta / \alpha \approx 2.44$. 
   It is found that the value of $\tau$ so obtained is distinctly different from the cluster size distribution exponent 
   $\tau$ = 2+5/91 $\approx$ 2.05 of the ordinary percolation problem in two dimensions.

\begin{figure}[t]
\begin {center}
\includegraphics[width=6.0cm]{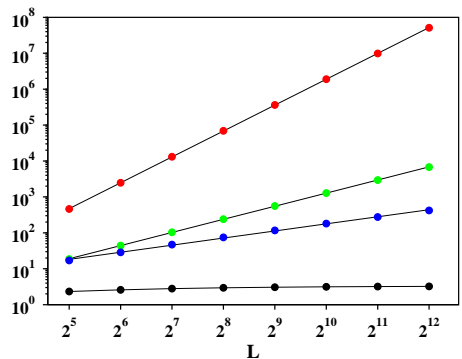}
\end {center}
\caption{
The average values of the sizes of the islands (without the infinite A cluster) have been plotted here
on the double logarithmic scale. The average size $\langle s(L) \rangle$ (black)
of the islands, does not increase against the system size $L$ and tends to 
saturate on increasing $L$. The other moments $\langle s^2(L) \rangle$ (blue), 
$\langle s_{max}(L) \rangle$ (green), and $\langle s_{max}^2(L) \rangle$ (red) 
vary as $L^{0.656}, L^{1.214}, L^{2.393}$ respectively.
}
\label {FIG04}
\end{figure}

      We have repeated the same exercise with a fixed linear probability gradient $g = 2^{-13}$ as well. On a 
   square lattice of rectangular shape, the probability gradient is applied along the $x$-axis of length
   $L_x = 8192$ and the three different values of the width $L_y$ = 512, 2048, and 8192 have been used. 
   In all cases the value of $p = p_c = 0.5927460507921$ \cite {Jacobsen} has been set exactly at i
   $x = L_x/2$. In Fig. \ref {FIG03}(a) we have shown the plot of the binned probability distribution data 
   $D(s,L_y)$ against the island size $s$ for these three system sizes. It is seen that even without any
   scale transformation of the coordinate axes, the data for the three system sizes collapse on top of 
   one another. A direct measurement of the slopes of the three curves in the most linear intermediate 
   region gives the estimate for the island size distribution exponent $\tau = 2.45 \pm 0.01$.

      The same simulation has been executed using an even smaller linear probability gradient $g = 2^{-17}$ 
   and for the same three rectangular system sizes. However, on plotting the $D(s,L_y)$ against $s$ data 
   in Fig. \ref {FIG03}(b), we find that there are two regions in each curve. In the regime of small 
   island sizes, the slopes of the curves are quite different and nearly 1.88. On the other hand for 
   large $s$ regime, the slopes are the same as we obtained before, i.e., 2.45. Finally, we present the
   same data for $g = 2^{-21}$ in Fig. \ref {FIG03}(c). Other than a bulge at the tail end of the
   distributions, the intermediate part is quite large and straight and has a common slope which is
   approximately 2.04, which is close to the value of the cluster size exponent 2.05 of ordinary 
   percolation in two dimensions.

      We try to explain the above three results in the following way. When we make the gradient $g$ very
   small if we could have make the lattice also large, like $L_x = 1/g$ then the sample of the island sizes
   used for estimating the probability distribution would have picked up island of all sizes, small, medium and large. 
   Instead, because of limitation of our computational resources, we have used a fixed value for $L_x = 8192$.
   Therefore, we have collected the island size data out of a box of fixed width centered around $p_c$.
   Smaller the value of the gradient $g$, the collected sample of island sizes is more close to $p_c$.
   This explains the crossover in the probability distribution of the island sizes where we get the single 
   exponent $\tau \approx 2.45$ when $L_x \sim 1/g$; the value of $\tau$ being close to the ordinary percolation 
   exponent 2.05 when $L_x << 1/g$ and simultaneously both values of $\tau$ for the intermediate values of $g$.
   
\begin{figure}[t]
\begin {center}
\includegraphics[width=6.0cm]{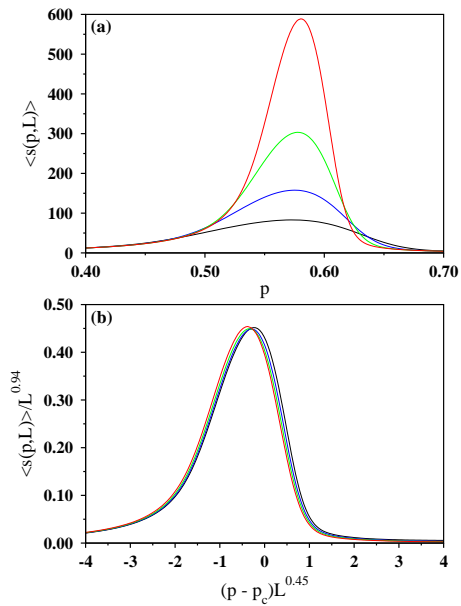}
\end {center}
\caption{
(a) The average island size $\langle s(p,L) \rangle$ has been plotted against $p$
    for $L$ = 256 (black), 512 (blue), 1024 (green), and 2048 (red). 
(b) First, the $x$ axis has been transformed to $p \rightarrow (p-p_c)$ and then both 
    $x$ and $y$ axes are scaled by the factors $L^{0.45}$ and $L^{0.94}$ respectively. 
}
\label {FIG05}
\end{figure}

      For the square lattices of square shapes of sizes $L \times L$, 
   we also identify the largest island and denote its size by $s_{max}$. We have estimated
   the average value $\langle s(L) \rangle$  of the sizes of the islands and the average size
   $\langle s_{max}(L) \rangle$ of the largest island over all independent configurations. In Fig. 
   \ref {FIG04}, we have plotted
   $\langle s(L) \rangle$ and $\langle s^2(L) \rangle$ against the system size. On increasing $L$, the 
   value of $\langle s(L) \rangle$ slightly increases initially but then tends to saturate to a constant value as
   $L$ becomes large. On the other hand $\langle s^2(L) \rangle$ nicely fits to a power law growth 
   $\langle s^2(L) \rangle \sim L^{0.656}$. Similar averages of the size of the largest cluster also fit
   to the power laws: $\langle s_{max}(L) \rangle \sim L^{1.214}$ and $\langle s^2_{max}(L) \rangle \sim L^{2.393}$
   respectively.

\begin{figure}[t]
\begin {center}
\includegraphics[width=7.0cm]{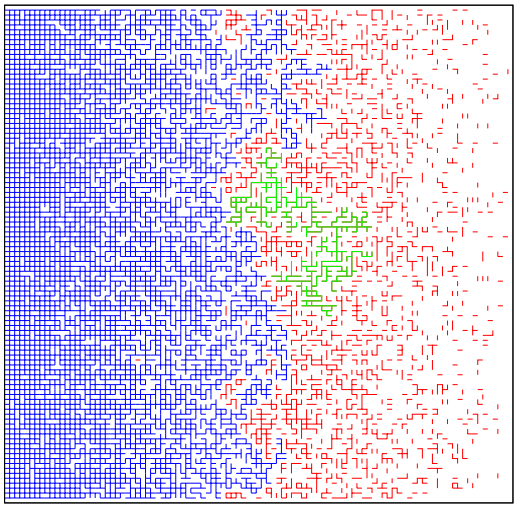}
\end {center}
\caption{
On a square lattice of size $L$ = 100, the upward vertical bond and the right horizontal bond
from each site $(x,y)$ have been occupied with probability $p(x) = 1 - x/L$ i.e., with gradient 
$g = 1/L$. The infinite A cluster has 7635 occupied bonds (blue), 
the largest island has 358 bonds (green) and all other islands has a total of 2339 
bonds (red).
}
\label {FIG06}
\end{figure}

      Looking at the Fig. \ref {FIG01} it is quite clear that the island sizes also depend
   on the position of the island. In other words, we expect that the average island size
   also depends on the occupation probability gradient set in the system. Therefore, we have
   defined that an occupied site at $x$ is a part of the island of size $\langle s(x,L) \rangle$ on the
   average. Since the probability $p(x)$ of occupation is an explicit function of $x$, we
   define the quantity $\langle s(p,L) \rangle$ instead. We have shown in Fig. \ref {FIG05}(a)
   the plot of $\langle s(p,L) \rangle$ against $p$ for the four different system sizes.
   Each plot has a peak which increases sharply as the system size increases. Each curve is asymmetric,
   i.e., on increasing $p$ it grows slowly in comparison, but beyond the peak it falls off faster. 
   Expectedly, the value of $p = p_{max}$ corresponding to the peak value slowly approaches to the
   percolation threshold $p_c$ as the system size $L$ increases. In Fig. \ref {FIG05}(b)
   we have performed a system size dependent scaling of the same data and plotted 
   $\langle s(p,L) \rangle / L^{0.94}$ against $(p - p_c)L^{0.45}$. The data collapse is found
   to be quite good. 

\section{Islands in gradient bond percolation}

      The problem of gradient percolation can also be studied in terms of lattice bonds,
   as in the ordinary bond percolation model. To the best of our knowledge, the model
   of gradient bond percolation has not been studied yet in the literature. The percolation
   threshold for bond percolation has been known to be exactly 1/2. Because of this,
   the set of occupied and vacant bonds are symmetric about the percolation point and we expect 
   to get better statistics compared to the gradient site percolation.

\begin{figure}[t]
\begin {center}
\includegraphics[width=6.0cm]{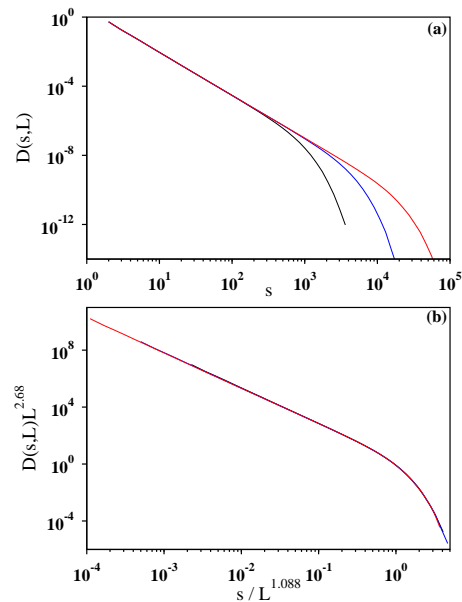}
\end {center}
\caption{For the gradient bond percolation:
(a) The probability distribution $D(s,L)$ of the island sizes have been plotted
    against the island size $s$ for square lattices of size $L \times L$. The system
    size dependent linear probability gradient $g = 1/L$ have been used for $L$ = 512 
    (black), 2048 (green), and 8192 (red).
(b) The same data have been replotted after scaling the $x$ and $y$ axes using 
    the scale factors $L^{1.088}$ and $L^{2.68}$ respectively. The data collapse
    appears to be excellent implying the island size distribution exponent 
    $\tau = 2.68/1.088 \approx 2.46$.}
\label {FIG07}
\end{figure}

      The probability gradient $g = 1/L$ has been applied along the $x$-axis as before. Therefore, 
   the position dependent bond occupation probability is $p(x)= 1-x/L$.
   Here, both the upward vertical bond and the right horizontal bond from $(x,y)$ have been occupied
   with probability $p(x)$. One gets a gradient bond percolation configuration when all bonds 
   of the lattice of size $L \times L$ are either occupied with probability $p(x)$ or left vacant
   with probability $1-p(x)$. For our island size 
   calculation, the equivalent site configuration has been created by occupying the end sites 
   of every occupied bond. As in the previous case, the burning algorithm has been applied to
   the resulting site configuration to estimate the island sizes, ignoring the infinite A cluster.
   The entire calculation has been repeated over a large number of independent configurations to obtain
   the probability distribution of island sizes. A typical gradient bond percolation configuration
   has been shown in Fig. \ref{FIG06}.
   
      The data for the probability distributions of gradient bond percolation have been plotted 
   in Fig. \ref {FIG07}(a). The above mentioned linear probability gradient have been applied 
   to three system sizes, namely, $L$ = 512, 2048, and 8192. Distributions $D(s,L)$ against $s$
   plots get separated from one another at their tail ends. However, in Fig. \ref {FIG07}(b),
   an excellent data collapse of the three curves were possible by scaling the axes using factors
   which are different powers of $L$. The best values of the scaling exponents have been estimated
   to be 1.088 and 2.68 for the $x$ and $y$ axes respectively, giving the value of the exponent 
   $\tau = 2.68/1.088 \approx 2.46$.

      The same probability distribution has be studied for a rectangular system of sizes $L_x$ = 
   8192 and $L_y$ = 512, 2048, and 8192 with fixed gradients $g = 2^{-13}, 2^{-17}$, and $2^{-21}$.
   A similar crossover between $\tau$ values 2.45 and 2.05 as reported in \ref {FIG03} has been 
   observed for the gradient bond percolation as well. Further, very similar results for the 
   variation of the average values of the island sizes as reported in \ref {FIG04} and \ref {FIG05}
   have been obtained. We therefore chose not to exhibit the similar plots here.

\begin{figure}[t]
\begin {center}
\includegraphics[width=7.0cm]{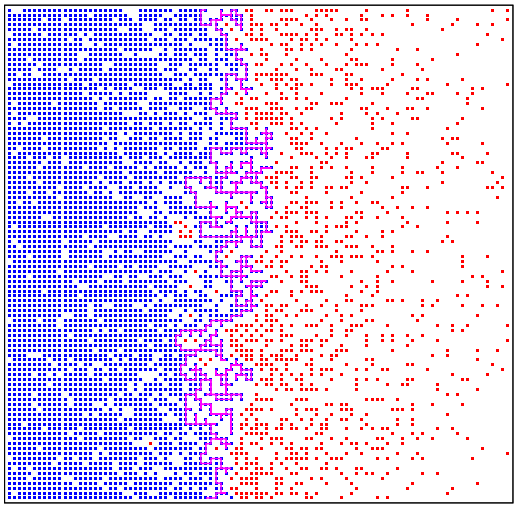}
\end {center}
\caption{
On an infinite strip of width $L$ = 100 placed along the $x$-axis, particles of an 
initial density profile with sharp interface starts diffusing. After a time of $T$ = 
128, the infinite A cluster has been shown by the 3853 blue dots, all finite islands by 
a total of 1147 red dots and the interface has been marked by the perimeter generating 
walk of 705 steps shown in magenta color.
}
\label {FIG08}
\end{figure}

\section{Islands near the interface of diffusion front}

      We have also studied the density gradient of a system of diffusing 
   particles. The diffusive dynamics starts from an initial configuration of particles 
   which is a fully occupied half lattice adjacent to a completely empty half lattice. 
   A square lattice in the form of an infinite strip of width $L$ parallel to the $x$ 
   axis has been used as the substrate. Initially, each site with $x$ coordinate in the 
   range $-\infty < x \le 0$ is occupied by a single particle and rest of the sites have
   been kept vacant. The average density $\rho(x,T)$ of particles along the vertical 
   line at $x$ is a function of both the position $x$ and the time $T$. Therefore, as per 
   construction, initially at time $T$ = 0, the density profile is a step function at 
   $x$ = 0, i.e., $\rho(x,0) = 1$ for $-\infty < x \le 0$ and 0 for $x > 0$. 

\begin{figure}[t]
\begin {center}
\includegraphics[width=6.0cm]{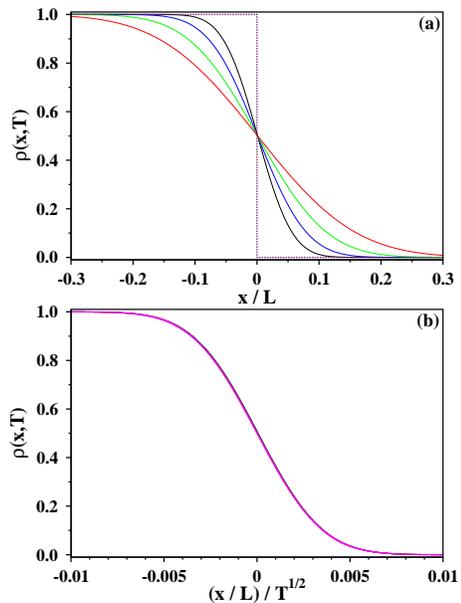}
\end {center}
\caption{
(a) Average particle density $\rho(x,T)$ has been plotted against the scaled coordinate $x/L$ 
    for the system size $L$ = 512 at different instants of time, e.g., $T$ = 0 (dashed 
    line), 256 (black), 512 (blue), 1024 (green), and 2048 (red).
(b) The same data have been replotted after a further scaling of the $x$-axis by a factor 
    $T^{1/2}$. In addition, we have plotted half of the complimentary error function 
    $(1/2)erfc(x/256)$ (magenta) to obtain an excellent collapse of the four scaled
    data sets and the function.
}
\label {FIG09}
\end{figure}

      The diffusion dynamics takes place in terms of an infinite sequence of particle hops.
   In a typical hop a particle selects randomly one of its neighboring sites with uniform 
   probability. The particle moves to that site if and only if the neighboring site is vacant 
   leaving its original site vacant. Clearly, the subset of particles which can at all make 
   successful hops are those which have at least one vacant neighbor. These particles are 
   referred as the `active particles' and their occupied lattice sites are called the `active 
   sites'. A lattice bond which has one occupied and one vacant site at its two ends must be 
   located on the perimeter of an island or a lake. Since the diffusion dynamics is limited 
   to these perimeter bonds only, we refer them as the `active bonds'. Therefore, an elementary 
   move consists of transferring a particle from one end of the active bond to the other end 
   and is equivalent to interchanging occupied and vacant status of the two end sites of the
   active bond. The diffusion dynamics constitutes the never ending sequence of such elementary 
   moves. 

      We have defined that one unit of time $T$ is completed when each active bond is updated
   once on the average by the random sequential updating procedure.

      The algorithm used to study this dynamics has been described in the following. At any intermediate 
   stage, the list of all active bonds are stored in an array, called `List'. For every
   active bond $i$ in the List its corresponding lattice 
   bond is assigned the number $i$. These numbers are stored in the array `Bondnumber'. 

\begin{figure}[t]
\begin {center}
\includegraphics[width=6.0cm]{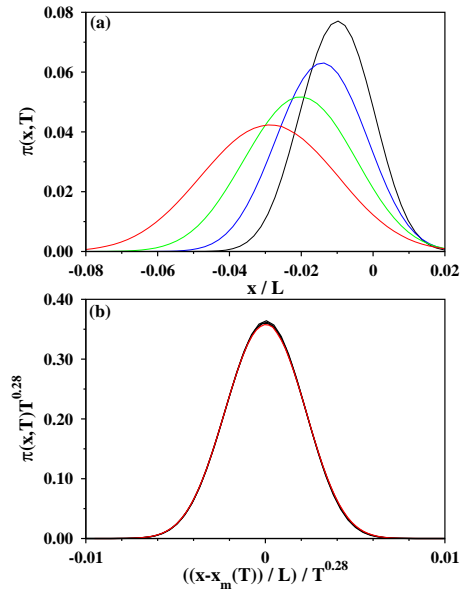}
\end {center}
\caption{
(a) The interface of the infinite A cluster of diffusing particles flattens with time. The 
    density $\pi(x,T)$ of the occupied sites on the interface has been plotted against the 
    scaled coordinate $x/L$ at different time instants of time, e.g., $T$ = 256 (black), 
    512 (blue), 1024 (green), and 2048 (red). Each curve fits excellent to a Gaussian and the 
    location $x_m(T)$ of the maximum has been obtained from the fit. As time passes the 
    location of the maximum drifts to the left, i.e., towards the high density region.
(b) The same data has been replotted, first by shifting an amount $x_m(T)$ and then scaling 
    both the axes by the same time dependent factor $T^{0.28}$ which exhibits nice collapse 
    of the data.
}
\label {FIG10}
\end{figure}

      An elementary move of the diffusive dynamics consists of selecting one of these active 
   bonds `$i$' randomly, with uniform probability. On the square lattice, the end sites of $i$ 
   are the meeting points of six other neighboring bonds, and in general some of them are 
   active and rest are inactive. The occupied and vacant status of the end sites of the bond 
   $i$ are now interchanged. As a result, the status of the bond $i$ remains unchanged but the 
   active / inactive status of the six neighboring bonds of $i$ are reversed. The neighboring 
   bonds which were active prior to the elementary move are now inactive and they are removed 
   from the List array and its length is shortened accordingly. The numbers associated with the
   corresponding bonds in the Bondbumber array are made zero. On the other hand, the bonds which 
   were inactive before the elementary move have now become active and therefore they are included 
   in the List array, its length is increased, and their bond numbers are stored in the Bondnumber 
   array.

      A snapshot of the particle configuration on an infinite strip placed along the $x$-axis 
   and of width $L = 100$ has been shown in Fig. \ref {FIG08}. Particles have been marked by the 
   blue dots. At time $T = 128$ many islands of occupied sites are visible within the initially 
   vacant half strip. Similarly, many lakes of vacant sites are visible within the initially 
   occupied half strip. Using the algorithm of `perimeter generating walk' \cite {Ziff} the 
   external hull of the infinite `A' cluster has been marked.

\begin{figure}[t]
\begin {center}
\includegraphics[width=6.0cm]{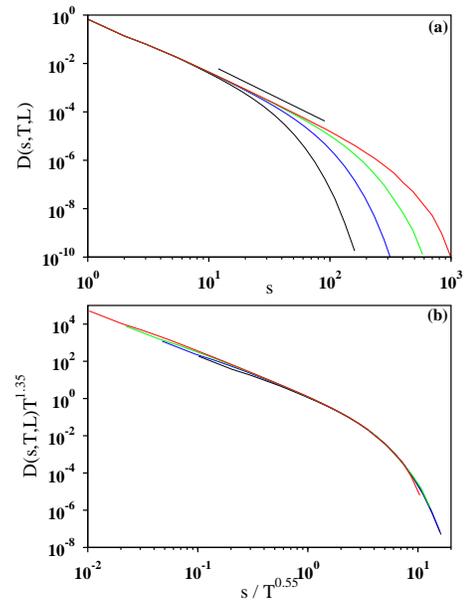}
\end {center}
\caption{
(a) The probability distribution $D(s,T,L)$ of island sizes at time $T$ has been plotted 
    against the island size $s$ of diffusing particles on an infinite strip of width $L$ = 
    1024. Four sets of data have been shown e.g., for $T$ = 64 (black), 256 (blue), 1024 
    (green), 4096 (red). A direct estimation of the slope of the curve for $T=4096$ at its 
    longest linear region yields an average value $\tau \approx$ 2.45, indicated by the 
    shifted straight line. 
(b) The same data have been replotted after scaling the axes using the time dependent factors 
    $T^{0.55}$ and $T^{1.35}$ for the $x$ and $y$ axes respectively. This analysis also gives 
    the same value of $\tau = 1.35/0.55 \approx$ 2.45.
}
\label {FIG11}
\end{figure}

      The density $\rho(x,T)$ of particles has been estimated as a function of $x$ and at different 
   instants of time $T$. The density has been averaged over all sites on the vertical
   line form $y = 1$ to $L$ and therefore it is not dependent on the $y$ coordinates. 
   In Fig. \ref {FIG09}(a) we have plotted the density $\rho(x,T)$ for the strip width $L$ = 512 against $x/L$ for the five 
   different values of $T$, namely, $T$ = 0, 200, 400, 800, and 1600. After the 
   initial step function, the density profile systematically flattens as the time proceeds.
   In Fig. \ref {FIG09}(b) we have replotted the same data but scaled the $x/L$ axis by $T^{1/2}$ to obtain
   the best collapse of the data. Further, in the same figure we have plotted half of the 
   complimentary error function $(1/2)\verb!erfc!(x/256)$ using the magenta color to obtain an excellent 
   data collapse. 

      As the density profile flattens with time, so also the interface of the infinite A 
   cluster. Like gradient percolation we have defined the interface of the infinite A cluster 
   as the set of occupied sites which have at least one vacant site in the neighboring, or 
   next neighboring sites that is connected via at least one path of B atoms to $x = +\infty$.
   At an arbitrary intermediate time, we have marked the occupied sites on the
   interface using the perimeter generating walk \cite {Ziff}. The density of these interface sites
   is denoted by $\pi(x,T)$. Initially, it is a delta function at $x = 0$. As time proceeds
   the density profile of the interface gradually expands in width and at the same time its height becomes shorter.
   However, the density profile fits to the Gaussian very well at all times and its peak position
   drifts slowly along the negative $x$-axis. In Fig. \ref {FIG10}(a) we have plotted the
   interface density profile for four different values of $T$, namely, for $T$ = 256, 
   512, 1024 and 2048. For each curve, we have denoted the peak position by 
   $x_m(T)$ and find that it shifts with time as: $x_m(T) \propto T^{0.509}$. In Fig. \ref {FIG10}(b) we have
   first shifted the $x$-axis by an amount equal to $x_m(T)$ so that the peak positions are
   on the same vertical line. Then we have tried to scale the $x$ and $y$ axes using a tunable power
   of time $T$ for a possible data collapse. It is found that a plot of $\pi(x,T)T^{0.28}$ against 
   $(x - x_m(T))/T^{0.28}$ gives the excellent data collapse of all four curves.

      Finally, the island size distribution $D(s,T,L)$ has been estimated for different values of
   time $T$ and system width $L$. In Fig. \ref {FIG11}(a) we have plotted the distribution
   $D(s,T,L)$ against island size $s$ at four different time instants, namely, $T$ = 64, 256, 1024
   and, 4096 for $L$ = 1024. These four curves are observed to nearly coincide up to the island size $\sim$10,
   but they separate out from one another for larger sizes. A scaled version of these
   curves have been plotted in Fig. \ref {FIG11}(b). In this figure, the $x$ and $y$ axes 
   have been scaled by the factors $T^{0.55}$ and $T^{1.35}$ respectively. Collapse
   of the data appear to be quite good. This implies that the value of the power law exponent associated with this
   distribution is $\tau = 1.35/0.55 \approx 2.45$.

\begin{figure}[t]
\begin {center}
\includegraphics[width=6.0cm]{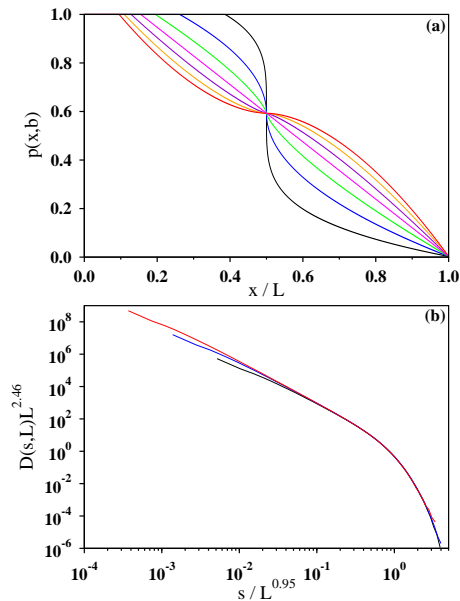}
\end {center}
\caption{
(a) Nonlinear occupation probability $p(x,b)$ has been plotted against the scaled $x/L$
    coordinate for the values of the power $b$ = 0.25 (black), 0.50 (blue), 0.75 (green),
    1.00 (magenta), 1.25 (maroon), 1.50 (orange), and 1.75 (red). A value of $L$ = 4096 has 
    been used for this plot.
(b) Finite size scaling analysis for the island size probability distribution using the 
    nonlinear percolation occupation probability with power $b=0.75$ for $L$ = 256 (black), 
    1024 (blue), and 4096 (red).
}
\label {FIG12}
\end{figure}

\section{Nonlinear gradient percolation}

      Finally, following the work of Gastner and Oborny \cite {Gastner}, we have studied the probability 
   distribution of the island sizes in a gradient percolation problem where the percolation occupation 
   probability $p(x)$ is a nonlinear function of $x$. More precisely, if the value of the percolation 
   threshold $p_c$ occurs at $x_c$ then in this case the percolation probability varies with the $x$ coordinate
   as $|p(x) - p_c| = a|x|^b$ where $a$ and $b$ are the tunable parameters. 

      In our simulation algorithm on the $L \times L$ square lattice we have assumed that $p = p_c$ occurs at $x = L/2$.
   The value of $a$ has been kept fixed at 0.59 where as seven different values of the parameter $b$ have been 
   studied (Fig. \ref {FIG12}). A large number of independent percolation configurations have been generated with 
   this prescription and
   the statistics of island sizes have been collected as before. Using the Eqn. (1) again, a similar scaling analysis 
   has been performed and the scaling exponents $\alpha(b)$, $\beta(b)$, and $\tau(b)$ have been found for every 
   value of $b$. These values have been tabulated in the Table 1.

\begin{figure}[t]
\begin {center}
\includegraphics[width=6.0cm]{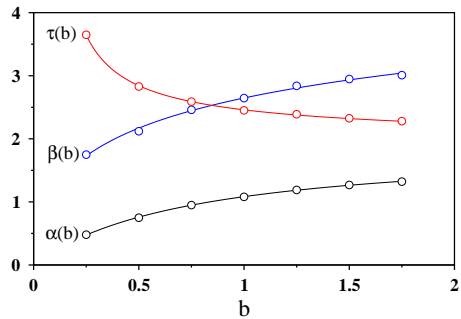}
\end {center}
\caption{
The best estimates of the finite size scaling exponents $\alpha(b)$ and $\beta(b)$
and the island size distribution exponent $\tau(b)$ have been plotted against the 
nonlinear power $b$ for seven different values of $b$. The data for $\alpha(b)$ vs. 
$b$ (black) have been best fitted by Eqn. 2 where as $\beta(b)$ vs.  $b$ (blue) and 
$\tau(b)$ vs. $b$ (red) have been best fitted by Eqn. 3. The data points have been 
shown by the circles where as the fitted curves have been shown by the continuous lines.
}
\label {FIG13}
\end{figure}

      In Fig. \ref {FIG13} we have shown the plots of $\alpha(b)$, $\beta(b)$ and $\tau(b)$ against $b$ whose values 
   have been listed in table 1. We try to fit their variations using some functional forms. However, without any theoretical
   argument we try with different functional forms similar to the Eqn. (3) in the paper \cite {Gastner}. We find that
   indeed a similar functional form exists and the $\alpha(b)$ vs. $b$ data have been fitted nicely by the equation:
\begin {equation}
\alpha(b) = b / (A_1b + A_2).
\end {equation}
   where $A_1$ = 0.528, $A_2$ = 0.395. It may be noted that the ratio $A_1 / A_2 \approx 1.337$ which is very close to 
   the correlation length exponent $\nu = 4/3$ of the ordinary two dimensional percolation. On the other hand, the values of 
   $\beta(b)$ vs. $b$ and $\tau(b)$ vs. $b$ have been fitted by the equations:
\begin {center}
\begin{tabular}{cc}
$\beta(b) = \sqrt{b} / (B_1\sqrt{b} + B_2)$  & and \\
$\tau(b)  = \sqrt{b} / (T_1\sqrt{b} + T_2)$. &\hspace*{3.0cm}(3)\\
\end {tabular}
\end {center}
   where $B_1$ = 0.177, $B_2$ = 0.200 and $T_1$ = 0.539, $T_2$ = -0.132.

   We thank the anonymous reviewer who pointed out that a negative value of $T_2$ would imply $\tau(b)$ to be
   infinite when $b \approx 0.06$. We believe that such a small non-zero value of $b$ is due to the 
   numerical uncertainties. Truly, the value of $\tau$ should be infinite for $b=0$, since for this value the
   probability function $p(x)$ is discontinuous as a step function. The islands occur only in the $x > L/2$ region
   where the value of $p(x) < p_c$ and therefore all islands in the sub-critical region are the small size clusters. 
   Like the ordinary percolation the probability distribution of these clusters should have an exponential 
   tail which is effectively equivalent to a power law tail with $\tau = \infty$.

\begin{table}[t]
\begin{tabular}{cccc} \hline
$b$   & {\hspace*{1.0cm}}$\alpha(b)$ & {\hspace*{1.0cm}} $\beta(b)$ & {\hspace*{1.0cm}}$\tau(b)$ \\ \hline
1.75  & {\hspace*{1.0cm}}1.321    & {\hspace*{1.0cm}} 3.01    & {\hspace*{1.0cm}}2.28   \\
1.50  & {\hspace*{1.0cm}}1.268    & {\hspace*{1.0cm}} 2.948   & {\hspace*{1.0cm}}2.325  \\
1.25  & {\hspace*{1.0cm}}1.19     & {\hspace*{1.0cm}} 2.84    & {\hspace*{1.0cm}}2.39   \\
1.00  & {\hspace*{1.0cm}}1.08     & {\hspace*{1.0cm}} 2.645   & {\hspace*{1.0cm}}2.45   \\
0.75  & {\hspace*{1.0cm}}0.95     & {\hspace*{1.0cm}} 2.46    & {\hspace*{1.0cm}}2.59   \\
0.50  & {\hspace*{1.0cm}}0.75     & {\hspace*{1.0cm}} 2.12    & {\hspace*{1.0cm}}2.83   \\
0.25  & {\hspace*{1.0cm}}0.50     & {\hspace*{1.0cm}} 1.90    & {\hspace*{1.0cm}}3.80   \\ \hline
\end {tabular}
\caption{Values of the island size distribution exponent $\tau(b)$ as well as the scaling exponents 
         $\alpha(b)$ and $\beta(b)$ for different values of $b$ in the nonlinear gradient percolation model.}
\label {TAB01}
\end{table}

\section{Summary}

      To summarize, we have revisited the well known problem of gradient percolation. In this problem, the
   sites of a regular lattice are occupied with a probability $p(x)$ that is a function of the $x$ 
   coordinate of the site. In the sections II and III, this function has a linear gradient with a constant 
   negative slope $g$ along the $x$-axis where we have done the calculation using the two methods, namely 
   (i) the gradient site percolation, and (ii) the gradient bond percolation. The model has been simulated 
   on a square lattice of rectangular shape whose width $L_x$ may be a maximum of $1/g$ and the height $L_y$. 
   The islands are the clusters of occupied sites within the infinite vacant cluster and the lakes are the 
   clusters of vacant sites within the infinite occupied cluster. The probability distribution of the island 
   sizes have been studied and found to follow a power law distribution with an exponent $\tau$. For $L_x \sim 1/g$
   the value of $\tau \approx 2.45$ is obtained which is distinctly different from the value of the same
   exponent of the ordinary two dimensional percolation problem. However, when $L_x << 1/g$ one retrieves 
   back the percolation cluster size exponent $\tau = 2+5/91 \approx 2.05$. Further in section IV, we have 
   studied the same distribution for a system of randomly diffusing particles starting from a dense region. 
   To the best of the accuracy of our simulations and analysis of the statistical data, we find very strong 
   indications that the value of the exponent $\tau \approx 2.45 \pm 0.02$ is indeed different from $5/2$ 
   though we cannot strictly rule out the small possibility that they may be the same. Finally, in section V
   we have studied the island size distribution again for the nonlinear gradient percolation following the
   work of Gastner and Oborny \cite{Gastner}. We have observed that the island size distribution exponent 
   $\tau$ is indeed an explicit function of the nonlinear parameter $b$.

      It's my great pleasure to congratulate Professor R. M. Ziff (Bob) on his 70th birth year. I wish him 
   a truly fabulous birth year and many more productive and successful years to come. I also thankfully 
   acknowledge Bob for many helpful discussions on this work and for the critical reading of the manuscript.

\end{document}